\documentclass[prb,superscriptaddress,amsfonts,twocolumn,showpacs,floatfix]{revtex4}
\usepackage{amsmath}
\usepackage{amsfonts}
\usepackage{amssymb}
\usepackage{amsmath}
\usepackage{txfonts}
\usepackage{tabularx}
\usepackage[dvips]{graphicx,color}

\begin{document}

\title{
Multiple superconducting transitions \\
in the Sr\boldmath$_3$Ru$_2$O$_7$ region of Sr$_3$Ru$_2$O$_7$\,--\,Sr$_2$RuO$_4$ eutectic crystals}
\author{S. Kittaka}
\affiliation{Department of Physics, Kyoto University, Kyoto 606-8502, Japan}
\author{S. Fusanobori}
\affiliation{Department of Physics, Kyoto University, Kyoto 606-8502, Japan}
\author{S. Yonezawa}
\affiliation{Department of Physics, Kyoto University, Kyoto 606-8502, Japan}
\author{H. Yaguchi}
\affiliation{Department of Physics, Kyoto University, Kyoto 606-8502, Japan}
\author{Y. Maeno}
\affiliation{Department of Physics, Kyoto University, Kyoto 606-8502, Japan}
\affiliation{International Innovation Center, Kyoto University, Kyoto 606-8501, Japan}
\author{R. Fittipaldi}
\affiliation{Department of Physics, Kyoto University, Kyoto 606-8502, Japan}
\affiliation{International Innovation Center, Kyoto University, Kyoto 606-8501, Japan}
\affiliation{CNR-INFM Regional Laboratory ``SuperMat" and Department of Physics University of Salerno, Baronissi (Sa), Italy }
\author{A. Vecchione}
\affiliation{Department of Physics, Kyoto University, Kyoto 606-8502, Japan}
\affiliation{International Innovation Center, Kyoto University, Kyoto 606-8501, Japan}
\affiliation{CNR-INFM Regional Laboratory ``SuperMat" and Department of Physics University of Salerno, Baronissi (Sa), Italy }

\date{\today}
\begin{abstract}
We report superconducting properties of Sr$_3$Ru$_2$O$_7$\,--\,Sr$_2$RuO$_4$ eutectic crystals, 
consisting of the spin-triplet superconductor Sr$_2$RuO$_4$ with a monolayer stacking of RuO$_2$ planes and 
the metamagnetic normal metal Sr$_3$Ru$_2$O$_7$ with a bilayer stacking. 
Although Sr$_3$Ru$_2$O$_7$ has not been reported to exhibit superconductivity so far,
our AC susceptibility measurements revealed multiple superconducting transitions occurring in the Sr$_3$Ru$_2$O$_7$ region of the eutectic crystals. 
The diamagnetic shielding essentially reached the full fraction at low AC fields parallel to the $c$ axis. 
However, both the shielding fraction and the onset temperature are easily suppressed by AC fields of larger than 0.1 mT-rms and no anomaly was observed in the specific heat.
Moreover, the critical field curves of these transitions have a positive curvature near zero fields, 
which is different from the upper critical field curve of the bulk Sr$_2$RuO$_4$.
These facts suggest that the superconductivity observed in the Sr$_3$Ru$_2$O$_7$ region is not a bulk property.
To explain these experimental results, we propose the scenario that 
stacking RuO$_2$ planes, the building block of superconducting Sr$_2$RuO$_4$, 
are contained in the Sr$_3$Ru$_2$O$_7$ region as stacking faults.

\end{abstract}
\maketitle
\section{Introduction}
The layered perovskite superconductor Sr$_2$RuO$_4$ ($T_\mathrm{c}=1.5$~K), 
isostructural to the high-$T_\mathrm{c}$ cuprate La$_{2-x}$Sr$_x$CuO$_4$, 
is now believed to be a spin-triplet superconductor with broken time-reversal symmetry
based on various experimental results.\cite{Maeno1994Nature,Mackenzie2003RMP,Ishida1998Nature,Luke1998Nature,Xia2006PRL}
After the discovery of superconductivity in Sr$_2$RuO$_4$, two types of eutectic solidification systems containing Sr$_2$RuO$_4$ have been grown: 
Sr$_2$RuO$_4$\,--\,Ru \cite{Maeno1998PRL} and Sr$_3$Ru$_2$O$_7$\,--\,Sr$_2$RuO$_4$ \cite{Fittipaldi2005JCG}
These eutectic systems are also interesting because they exhibit unusual superconducting features.

The Sr$_2$RuO$_4$\,--\,Ru eutectic system,\cite{Maeno1998PRL} in which lamellae of Ru metal are embedded in Sr$_2$RuO$_4$,
exhibits a large enhancement of $T_\mathrm{c}$. 
AC susceptibility measurements \cite{Maeno1998PRL} revealed a broad diamagnetic transition with an onset temperature as high as 3~K, 
which is twice higher than those of best-quality Sr$_2$RuO$_4$ single crystals. 
Therefore, this eutectic is referred to as the 3-K phase.
However, specific heat measurements \cite{Yaguchi2003PRB} revealed that 
the volume fraction of the superconductivity associated with the 3-K phase is very small. 
Measurements of the tunneling conductance between Sr$_2$RuO$_4$ and a single Ru lamella \cite{Kawamura2005JPSJ,Yaguchi2006JPSJ} support that 
the superconductivity with an enhanced $T_\mathrm{c}$ occurs in the boundaries between Sr$_2$RuO$_4$ and embedded Ru lamellae.

We have recently succeeded in growing another Sr$_2$RuO$_4$-based eutectic system \cite{Fittipaldi2005JCG}: Sr$_3$Ru$_2$O$_7$\,--\,Sr$_2$RuO$_4$. 
This eutectic system consists of the spin-triplet superconductor Sr$_2$RuO$_4$ with a monolayer stacking of RuO$_2$ planes and the
metamagnetic normal metal \cite{Perry2004PRL,Borzi2004PRL} Sr$_3$Ru$_2$O$_7$, which consists of a bilayer stacking. 
X-ray diffraction analyses of the Sr$_3$Ru$_2$O$_7$\,--\,Sr$_2$RuO$_4$ eutectic crystals indicated that 
the directions not only of the $c$ axis but also of the in-plane axes of Sr$_2$RuO$_4$ and Sr$_3$Ru$_2$O$_7$ are common in the eutectic crystals. \cite{Fittipaldi2005JCG}
The superconductivity observed in this eutectic crystal also exhibits interesting features.
AC susceptibility measurements \cite{Fittipaldi2005JCG} of an eutectic sample containing a number of Sr$_2$RuO$_4$ and Sr$_3$Ru$_2$O$_7$ domains 
revealed that a superconducting transition occurs at 1.43 K and the 
diamagnetic shielding fraction keeps increasing upon cooling well below $T_\mathrm{c}$. 
It was speculated that this additional diamagnetic signal was due to a proximity effect into Sr$_3$Ru$_2$O$_7$ from superconducting Sr$_2$RuO$_4$. \cite{Fittipaldi2005JCG}

Subsequently, a finite superconducting critical current in a Sr$_3$Ru$_2$O$_7$\,--\,Sr$_2$RuO$_4$ eutectic system 
containing many Sr$_2$RuO$_4$ and Sr$_3$Ru$_2$O$_7$ domains was observed by Hooper $et$ $al$. \cite{Hooper2006PRB} 
Their finding appears to indicate that Sr$_3$Ru$_2$O$_7$ domains are also superconducting. 
They suggested the possibility of a proximity effect in the Sr$_3$Ru$_2$O$_7$ regions with an unusually-long coherence length. 
In fact, the coherence length $\xi_\mathrm{N}$ in Sr$_3$Ru$_2$O$_7$ due to a proximity effect 
must be as long as the size of Sr$_3$Ru$_2$O$_7$ domains, a few hundred micro meters, 
if supercurrent flows across the Sr$_3$Ru$_2$O$_7$ regions.
However, the conventional coherence length of a proximity effect in a clean limit approximation yields 
$\xi_\mathrm{N} \thicksim$ 0.17 $\muup$m at 0.3 K.  
This value of the conventional coherence length is too short to account for the superconductivity in Sr$_3$Ru$_2$O$_7$. 

In the present study, 
we investigated the temperature dependence of AC susceptibility at various AC and DC fields, 
using Sr$_3$Ru$_2$O$_7$\,--\,Sr$_2$RuO$_4$ eutectic samples 
consisting of one Sr$_2$RuO$_4$ region and one Sr$_3$Ru$_2$O$_7$ region with a single boundary between them (e.~g. see the insets of Fig.~\ref{cut214}(a) and Fig.~\ref{AC}). 
Measurements of these samples revealed that the apparent superconducting volume fraction of the Sr$_3$Ru$_2$O$_7$\,--\,Sr$_2$RuO$_4$ eutectic sample was as large as 100\%. 
In order to test the proximity scenario, we performed similar measurements with the Sr$_3$Ru$_2$O$_7$ region cut from a eutectic crystal 
and, surprisingly, we also observed superconductivity with a very large apparent volume fraction. 
These results indicate that the superconductivity with a large apparent volume fraction occurs in the Sr$_3$Ru$_2$O$_7$ region and 
that its origin cannot be attributed to a proximity effect from the bulk Sr$_2$RuO$_4$ region. 
In addition, we did not observe any anomaly in the specific heat of the Sr$_3$Ru$_2$O$_7$ region.
Also, we calculated the temperature dependence of the AC susceptibility based on a multiple superconductor model (Scenario II in Sec.~\ref{disc}) and 
obtained calculated results which well match our experimental results.

\section{Experimental}
In this paper, we mainly present data which were obtained using a sample cut from a Sr$_3$Ru$_2$O$_7$\,--\,Sr$_2$RuO$_4$ eutectic crystal 
(batch No.~Cfv07 in Ref.~\onlinecite{Fittipaldi2005JCG}), grown with a floating-zone furnace. 
We carefully chose a Sr$_3$Ru$_2$O$_7$\,--\,Sr$_2$RuO$_4$ eutectic part 
which has only one boundary between Sr$_2$RuO$_4$ and Sr$_3$Ru$_2$O$_7$ (hereafter referred to as Sample~1). 
The size of Sample~1 was approximately $1.5\times 0.7\times 0.3\ \mathrm{mm^3}$. 
The inset of Fig.~\ref{cut214}(a) shows polarized light optical microscopy (PLOM) images of polished $ab$ planes of Sample~1. 
The darker area of the sample is the Sr$_2$RuO$_4$ region and the brighter area is the Sr$_3$Ru$_2$O$_7$ region, 
which was confirmed by a high resolution X-ray diffractometer \cite{Fittipaldi2005JCG} and energy dispersive X-ray (EDX) analysis. 
Sample~1 certainly consists of one bulk Sr$_2$RuO$_4$ region and one bulk Sr$_3$Ru$_2$O$_7$ region 
because the top and bottom surfaces have the same eutectic pattern. 
In order to check the reproducibility of experimental results, we performed measurements with more than ten eutectic samples 
(one of them is Sample~2 from the batch Cfv07, shown in the inset of Fig.~\ref{AC}(c)). 
Eutectic samples from different batches exhibit qualitatively the same behavior, too. 

We measured AC magnetic susceptibility $\chi_\mathrm{AC}=\chi^\prime+i \chi^{\prime \prime}$ by a mutual-inductance technique using a lock-in amplifier at various frequencies ranging from 19 Hz to 3011 Hz. 
The data shown below were all taken at 3011~Hz because the frequency dependence of $\chi_\mathrm{AC}$ was found to be insignificant.
The AC susceptibility was measured down to 0.3~K using a $^3$He cryostat with a 2-T magnet (Oxford Instruments), 
and down to 20 mK using a $^3$He-$^4$He dilution refrigerator (Cryoconcept) with an 11-T magnet (Oxford Instruments). 
The AC field $H_\mathrm{AC}$ was applied parallel to the $c$ axis or the $ab$ plane with a small coil (40~$\muup$T / mA), and 
the DC field $H_\mathrm{DC}$ was applied parallel to $H_\mathrm{AC}$. 
In this paper, we mainly report results under magnetic fields parallel to the $c$ axis.
When we measured $\chi_\mathrm{AC}$ in zero DC field, we used a high-permeability-metal shield to exclude the geomagnetic field of about 50~$\muup$T. 
The resultant residual field in this shield was estimated to be lower than 0.1~$\muup$T.

We also measured specific heat $c_p$ of eutectic samples by a thermal relaxation method with a commercial calorimeter (Quantum Design, PPMS) from 30 K to 0.8 K.
The Sr$_3$Ru$_2$O$_7$\,--\,Sr$_2$RuO$_4$ eutectic crystals were characterized by X-ray diffraction (XRD) with CuK$_{\alpha1}$ radiation and EDX analysis.

\section{Results}

\subsection*{AC susceptibility measurement}

\begin{figure}
\includegraphics[width=3.3in]{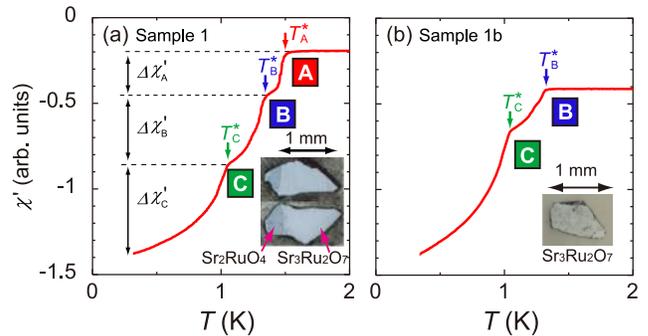}
\caption{(Color online) Temperature dependence of the real part of the AC susceptibility at $\mu_0H_\mathrm{AC}$ = 0.58 $\muup$T-rms and $\mu_0H_\mathrm{DC}$ = 0 T: 
(a) for Sample~1 (Sr$_3$Ru$_2$O$_7$\,--\,Sr$_2$RuO$_4$ eutectic); 
(b) for Sample~1b (i.\,e.\ the Sr$_3$Ru$_2$O$_7$ region cut from Sample~1).
Insets are PLOM images of the samples.}
\label{cut214}
\end{figure}

\begin{figure}
\includegraphics[width=3.3in]{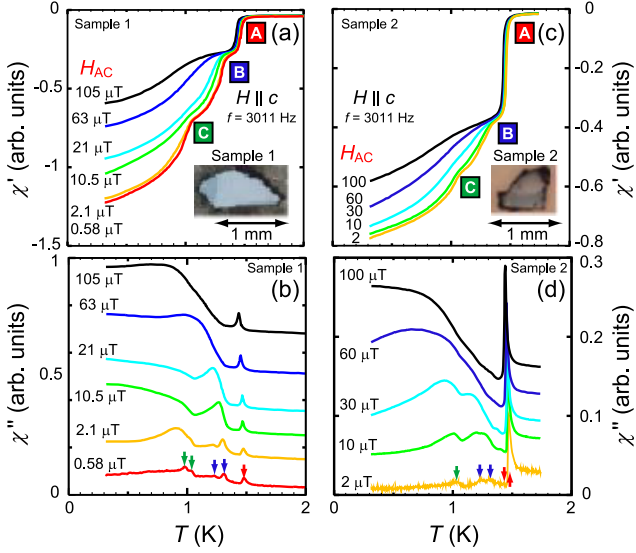}
\caption{(Color online) Temperature dependence of the AC susceptibility of Sr$_3$Ru$_2$O$_7$\,--\,Sr$_2$RuO$_4$ eutectic crystals in various AC magnetic fields 
($H_\mathrm{AC} \parallel c$, $\mu_0H_\mathrm{DC}=0$ T).
Figures~(a) and (b) represent the real and imaginary parts of $\chi_\mathrm{AC}$ of Sample~1, respectively.
Figures~(c) and (d) represent those of Sample~2.
Insets are PLOM images of the samples.
The numbers labeling the curves give the applied AC field amplitude $\mu_0H_\mathrm{AC}$ in $\muup$T-rms.}
\label{AC}
\end{figure}

Figure~\ref{cut214}(a) shows the temperature dependence of the AC susceptibility of Sample~1 (Sr$_3$Ru$_2$O$_7$\,--\,Sr$_2$RuO$_4$ 
eutectic crystal) 
under $\mu_0H_\mathrm{AC}$ = 0.58 $\muup$T-rms and $\mu_0H_\mathrm{DC}$ = 0 T with a high-permeability metal shield. 
In this measurement, we observed three steep changes of the diamagnetic signal in $\chi^\prime$, 
which hereafter we call transitions A, B, and C. 
The transition temperatures, 
defined as the onset temperatures of the transitions in $\chi^\prime$, 
were 1.48 K, 1.33 K, and 1.04 K, and are hereafter labeled $T^*_\mathrm{A}$, $T^*_\mathrm{B}$, and $T^*_\mathrm{C}$, respectively. 
Although more peaks were observed in $\chi^{\prime \prime}$ (marked with the arrows in Fig.~\ref{AC}(b)),
we mainly focus on the transitions A, B, and C. 
It was difficult to evaluate the shielding fraction accurately because of the large demagnetization factor of the sample.
By comparing the diamagnetic signal 
$\Delta \chi^\prime$ of Sample~1, which is equal to $\Delta \chi^\prime_\mathrm{A}+\Delta \chi^\prime_\mathrm{B}+\Delta \chi^\prime_\mathrm{C}$ as shown in Fig.~\ref{cut214}(a), with that of a pure Sr$_2$RuO$_4$ crystal 
with dimensions similar to that of Sample~1, we evaluated the apparent shielding fraction of Sample~1 to be approximately 100\% at 0.3~K. 
This large shielding fraction implies that 
the superconducting screening current flows not only in the Sr$_2$RuO$_4$ part of the sample but also in most of the Sr$_3$Ru$_2$O$_7$ region. 


In order to clarify whether or not this superconductivity is attributed to an unusual proximity effect from the boundary of the bulk Sr$_2$RuO$_4$, 
we completely removed the bulk Sr$_2$RuO$_4$ part from Sample~1, hereafter labeled Sample~1b.
The dimension of Sample~1b is approximately $1.0\times 0.6\times 0.15\ \mathrm{mm^3}$. 
The inset of Fig.~\ref{cut214}(b) is a PLOM image of Sample~1b.
As presented in Fig.~\ref{cut214}(b), two of the superconducting transitions were still observed in Sample~1b. 
The transition temperatures were 1.32 K and 1.04 K, 
well corresponding to $T^*_\mathrm{B}$ and $T^*_\mathrm{C}$ in Sample~1. 
The absence of transition~A in Sample~1b proves that transition~A originates from the bulk Sr$_2$RuO$_4$ part of Sample~1 
and that the transitions B and C occur in the Sr$_3$Ru$_2$O$_7$ region. 
The apparent shielding fraction of Sample~1b was estimated to be 90\% for $H \parallel c$ 
from the diamagnetic signal $\Delta \chi^\prime ( = \Delta \chi^\prime_\mathrm{B} + \Delta \chi^\prime_\mathrm{C})$.
In contrast, it was estimated to be less than 1\% for $H \parallel ab$ (not shown).
These facts indicate that the superconducting screening current mainly flows within the $ab$ planes.
From these measurements, we conclude that the Sr$_3$Ru$_2$O$_7$ region in the eutectic crystal has multiple superconducting transitions 
though pure Sr$_3$Ru$_2$O$_7$ has not been reported to become superconducting down to 20 mK (Ref. \onlinecite{Perry2004PRL}). 
Moreover, it is clear that the origin of the superconductivity observed in the Sr$_3$Ru$_2$O$_7$ region is not a proximity effect from the bulk Sr$_2$RuO$_4$ part of eutectic crystals across the boundary. 


We revealed that \textit{both} $T^*$ and $\Delta \chi^\prime$ of the transitions B and C are extremely sensitive to the amplitude of $H_\mathrm{AC}$ when $H_\mathrm{AC}$ is parallel to the $c$ axis. 
Figures~\ref{AC}(a) and (b) represent $\chi^\prime$ and $\chi^{\prime \prime}$ of Sample~1 in different AC magnetic fields. 
We normalized the obtained signals with respect to the strength of $H_\mathrm{AC}$. 
As shown in Fig.~\ref{AC}(a), $T^*_\mathrm{A}$ and $\Delta\chi^\prime_\mathrm{A}$ hardly depends on $H_\mathrm{AC}$ up to 100 $\muup$T-rms. 
In contrast, $\Delta\chi^\prime_\mathrm{B}$ and $\Delta\chi^\prime_\mathrm{C}$ are severely suppressed by $H_\mathrm{AC}$ of less than 100 $\muup$T-rms.
In addition, $T^*_\mathrm{B}$ and $T^*_\mathrm{C}$ are easily shifted toward lower temperatures with increasing AC field amplitude.
As represented in Figs.~\ref{AC}(c) and (d), we reproducibly observed these features in other samples.
However, when $H_\mathrm{AC}$ is applied parallel to the $ab$ plane,
$T^*$ and $\Delta \chi^\prime$ of the transitions B and C are not sensitive to $H_\mathrm{AC}$ of less than 100 $\muup$T-rms. 


\begin{figure}
\includegraphics[width=3.3in]{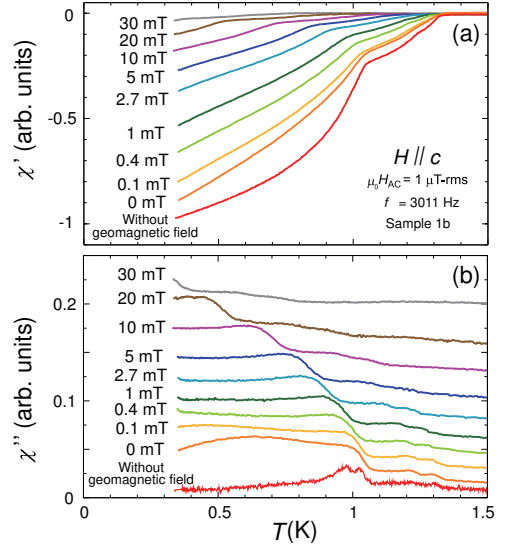}
\caption{(Color online) Temperature dependence of the (a) real and (b) imaginary parts of $\chi_\mathrm{AC}$ of Sample~1b in different DC fields. 
}
\label{DC1}
\end{figure}

\begin{figure}
\includegraphics[width=3.3in]{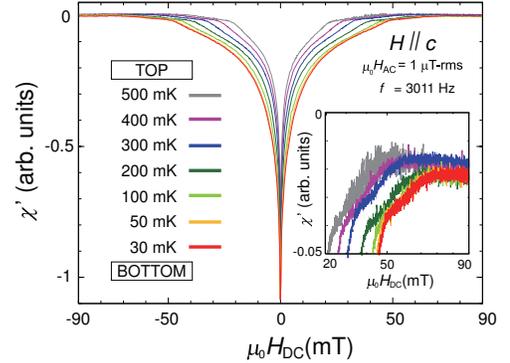}
\caption{(Color online) The DC field dependence of the real part of $\chi_\mathrm{AC}$ of Sample~1b.
The inset gives an enlarged view near onset. 
In order to take into account the residual fields in our equipment, which is estimated to be approximately 1 mT, 
we determined zero field as a field at which $\chi^\prime$ becomes minimum.
}
\label{DC2}
\end{figure}

\begin{figure}
\includegraphics[width=3in]{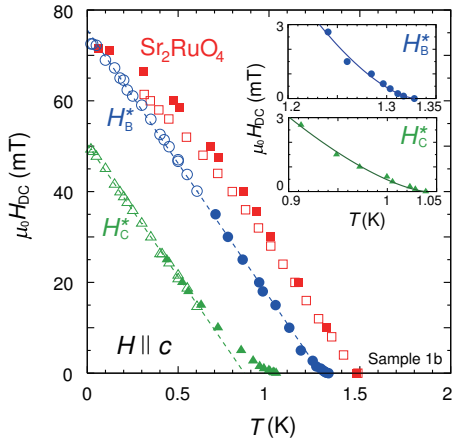}
\caption{(Color online) $H$\,--$T$ phase diagram of Sample~1b for $H_\mathrm{AC} \parallel H_\mathrm{DC} \parallel c$ determined from the AC susceptibility measurements.
The circles and triangles represent the critical fields of the transitions B and C, respectively.
The open symbols are obtained from field sweep data and the closed symbols from temperature sweep data. 
$H_\mathrm{c2}$ of bulk Sr$_2$RuO$_4$ from specific-heat data (Ref.~\onlinecite{Deguchi2004JPSJ}, closed squares) and 
from AC susceptibility measurements (Ref.~\onlinecite{Mao1999PRB}, open squares) are also plotted.
The insets show the low-field region below 3 mT.
The solid lines represent fits of the data close to $T^*_0\equiv T^*(H_\mathrm{DC}=0)$ using the function $\alpha(1-T/T^*_0)^n$.
The dashed lines present results of linear fittings to the data between 0 K and 0.7 $T^*_0$.}
\label{HTc}
\end{figure}

In order to obtain more information on the transitions B and C, we measured the DC field dependence of $\chi_\mathrm{AC}$ for Sample~1b.
Figure~\ref{DC1} shows the temperature dependence of $\chi^\prime$ and $\chi^{\prime \prime}$ at various DC fields and 
Fig.~\ref{DC2} presents the DC field dependence of $\chi^\prime$ at several temperatures for Sample~1b. 
In these measurements, 
we fixed the amplitude of $H_\mathrm{AC}$ to 1 $\muup$T-rms, and both $H_\mathrm{AC}$ and $H_\mathrm{DC}$ were applied parallel to the $c$ axis.
These measurements revealed that $T^*_\mathrm{B}$ and $T^*_\mathrm{C}$ are \textit{not} severely suppressed, but 
$\Delta \chi^\prime_\mathrm{B}$ and $\Delta \chi^\prime_\mathrm{C}$ are easily suppressed by $H_\mathrm{DC}$.

We obtained the $H$\,--$T$ phase diagram for $H \parallel c$, which is plotted in Fig.~\ref{HTc}.
Here, the critical fields of the transitions B and C are labeled $H_\mathrm{B}^*$ and $H_\mathrm{C}^*$, respectively, 
which are defined as the onset of $\chi^\prime$.
For comparison, we included the upper critical field $H_\mathrm{c2}$ of bulk Sr$_2$RuO$_4$ 
determined by specific heat measurements \cite{Deguchi2004JPSJ} and 
those determined by AC susceptibility measurements \cite{Mao1999PRB} in this figure.
The extrapolation of $\mu_0H_\mathrm{B}^*$ to $T=0$ yields 75 mT, which is nearly equal to $\mu_0H_\mathrm{c2}(T=0)$ of bulk Sr$_2$RuO$_4$.

However, temperature dependences of $H_\mathrm{B}^*$ and of $H_\mathrm{C}^*$ are qualitatively different from $H_\mathrm{c2}(T)$ of bulk Sr$_2$RuO$_4$.
Fitting the function $H_\mathrm{c2}(T)=\alpha(1-T/T_\mathrm{c})^n$ to the $H_{c2}$ data from specific heat measurements of bulk Sr$_2$RuO$_4$ yields $n=1.0$ for $H \parallel c$ near $H_\mathrm{DC}=0$, 
where $\alpha$ and $n$ are adjustable parameters. 
In contrast, both $H_\mathrm{B}^*$ and $H_\mathrm{C}^*$ exhibit 
temperature dependences with positive curvatures ($n=1.6$ for $H_\mathrm{B}^*$, $n=1.5$ for $H_\mathrm{C}^*$) 
near $H_\mathrm{DC}=0$ 
and then increase approximately linearly with decreasing temperature.
Such behavior suggests that the transitions B and C are of a similar origin, 
but different from the bulk superconducting transition.

\begin{figure}
\includegraphics[width=3in]{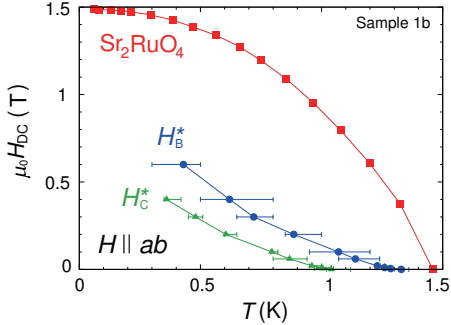}
\caption{(Color online) $H$\,--$T$ phase diagram of Sample~1b for $H_\mathrm{AC} \parallel H_\mathrm{DC} \parallel ab$.
The circle and triangular symbols represent the critical fields of the transitions B and C determined by the AC susceptibility measurements, respectively.
The upper critical fields $H_\mathrm{c2}$ of bulk Sr$_2$RuO$_4$ are obtained from AC susceptibility measurements (Ref.~\onlinecite{Mao1999PRB}, squares).
The solid lines are guides to the eye.
}
\label{HTab}
\end{figure}

We also constructed the $H$\,--$T$ phase diagram for $H \parallel ab$, as shown in Fig.~\ref{HTab}.
In this measurement, both $H_\mathrm{AC}$ of 20 $\muup$T-rms and $H_\mathrm{DC}$ were applied parallel to the $ab$ plane. 
Both the temperature dependence of $H^*_\mathrm{B}$ and of $H^*_\mathrm{C}$ 
exhibit a positive curvature near $H_\mathrm{DC}=0$, similar to those for $H \parallel c$. 
The critical field anisotropies $H_{\parallel ab}^* / H_{\parallel c}^*$ of the transitions B and C are approximately 13 at 0.3~K, 
which is somewhat smaller than that observed for bulk Sr$_2$RuO$_4$ (Ref.~\onlinecite{Deguchi2002JPSJ}: $H_{\mathrm{c2} \parallel ab} / H_{\mathrm{c2} \parallel c} \sim$ 20). 

\subsection*{Specific heat measurement}

\begin{figure}
\includegraphics[width=3in]{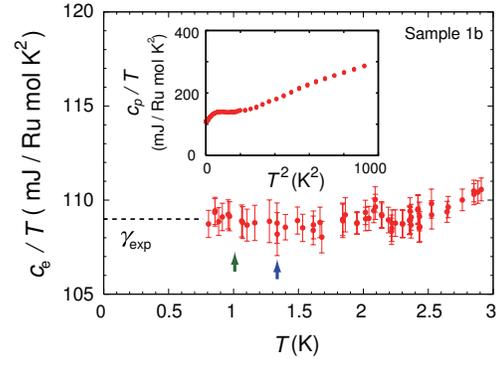}
\caption{(Color online) Temperature dependence of the electronic specific heat divided by temperature $c_\mathrm{e}/T$ of Sample~1b. 
The inset represents $c_p/T$ plotted against $T^2$.
The arrows mark $T^*_\mathrm{B}$ and $T^*_\mathrm{C}$ of this sample at zero field.}
\label{Cp}
\end{figure}

We measured the specific heat of Sample~1b (with a mass $m=0.472$ mg), which exhibits nearly full diamagnetic shielding in our AC susceptibility measurements for $H \parallel c$.
This specific heat measurement was performed in zero field, but the geomagnetic field and the residual field of the magnet ($\lesssim$ 1 mT) were not shielded.
The main panel of Fig.~\ref{Cp} shows the electronic specific heat divided by temperature for Sample~1b. 
There is no anomaly at $T^*_\mathrm{B}$ and $T^*_\mathrm{C}$. 
Therefore, we conclude that the actual volume fraction of the superconductivity observed in the apparent Sr$_3$Ru$_2$O$_7$ region is very small.

In order to obtain the electronic specific-heat coefficient $\gamma_\mathrm{exp}$ of Sample~1b and 
check the molar ratio $x$ of Sr$_2$RuO$_4$ contained in Sample~1b, 
we used an effective weight per Ru-mol $M_\mathrm{eff}$, which is defined as $M_\mathrm{eff}(x)=xM_\mathrm{214}+(1-x)M_\mathrm{327}/2$, 
where $M_\mathrm{214}$ and $M_\mathrm{327}$ are the molar weights of a formula unit of Sr$_2$RuO$_4$ and Sr$_3$Ru$_2$O$_7$, respectively.
We determined $\gamma_\mathrm{exp}$, which is obtained from the relation $\gamma_\mathrm{exp}(x)=(C_p/T)m/M_\mathrm{eff}(x)$ 
($C_p$ is the heat capacity of the sample), self-consistently by adjusting $x$ so that 
$\gamma_\mathrm{exp}(x)$ becomes equal to $x \gamma_{214} +(1-x) \gamma_{327}$. 
Here, $\gamma_{214}$ and $\gamma_{327}$ represents the electronic specific-heat coefficient of 
bulk Sr$_2$RuO$_4$ ($\gamma_{214}=38$ mJ/Ru-mol\,K$^2$ (Ref.~\onlinecite{Maeno1997JPSJ})) and 
Sr$_3$Ru$_2$O$_7$ ($\gamma_{327}=110$ mJ/Ru-mol\,K$^2$ (Ref.~\onlinecite{Ikeda2000PRB})), respectively.
As a result, we obtained $\gamma_\mathrm{exp} \sim$ 109 mJ/Ru-mol\,K$^2$ and $x=0.016 \pm 0.008$.
In addition, the overall temperature dependence 
of the total specific heat $c_p$ of Sample~1b presented in the inset of Fig.~\ref{Cp} is consistent with previous reports \cite{Ikeda2000PRB} for pure Sr$_3$Ru$_2$O$_7$.
These facts imply that the Sr$_3$Ru$_2$O$_7$ region of the eutectic crystals is almost the same as pure Sr$_3$Ru$_2$O$_7$.

\subsection*{PLOM, EDX, and XRD analyses}

\begin{figure}
\includegraphics[width=3in]{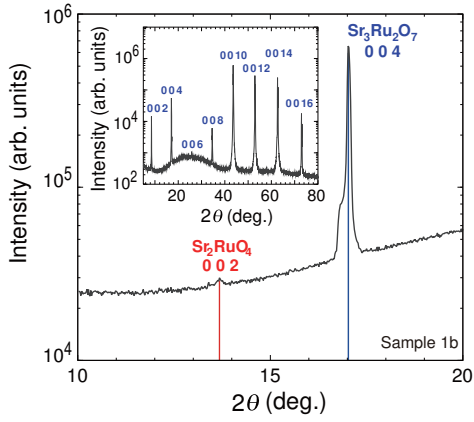}
\caption{(Color online) X-ray diffraction pattern for the $ab$ plane of Sample~1b.
Note that the vertical axis is in a logarithmic scale.}
\label{Xray}
\end{figure}

In order to characterize the sample in more details,
we took PLOM images, and performed elemental composition analysis with an EDX spectrometer and XRD analysis. 
From PLOM images and elemental composition analysis, we did not find Sr$_2$RuO$_4$ and 
the whole sample seemed to consist of Sr$_3$Ru$_2$O$_7$.
We note that we cannot rule out the presence of Sr$_2$RuO$_4$ parts with a size of less than about 1 $\muup$m, 
which is the experimental resolution limit of our instruments.
In the XRD pattern for the $ab$ plane of Sample~1b, as shown in Fig.~\ref{Xray}, 
a very weak $\langle 0 0 2 \rangle$ peak of Sr$_2$RuO$_4$ was detected in addition to strong Sr$_3$Ru$_2$O$_7$ peaks.
The observed peak intensity suggests that less than a few percent Sr$_2$RuO$_4$ is contained at least in the surface region of Sample~1b. 
This possible small content of Sr$_2$RuO$_4$ is consistent with the results of the specific heat measurement.

\section{Discussion}
\label{disc}

In order to discuss why superconductivity is observed in the Sr$_3$Ru$_2$O$_7$ region of eutectic crystals, 
we assumed two scenarios (Scenario~I and II) and calculated $\chi_\mathrm{AC}(T)$ for $H \parallel c$ using simplified models. 

\subsection*{Scenario I}

\begin{figure}
\includegraphics[width=2.9in]{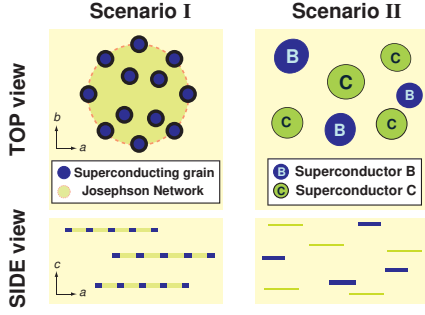}
\caption{(Color online) Scenarios for the superconductivity observed in the Sr$_3$Ru$_2$O$_7$ region of eutectic crystals are depicted.
The superconducting regions are probably distributed along the $ab$ planes because the shielding currents mainly flow within the $ab$ planes.
Such a layered arrangement is not taken into account in our simple model calculations.}
\label{model}
\end{figure}

First, we note that our results of $\chi_\mathrm{AC}(T)$ is somewhat similar to those of granular superconductors, 
in which Josephson-type weak links are formed among superconducting grains. 
For example, polycrystals of Tl$_2$Ba$_2$Ca$_2$Cu$_3$O$_y$, prepared by a suitable sintering process, consist of agglomerates of grains
whose typical size is approximately 5 $\muup$m. \cite{Yang1994PRB}
These grains are arranged randomly and connected strongly by nonstoichiometric interfacial materials. 
When such polycrystals are cooled below $T_\mathrm{c}$ of the grains, the grains first become superconducting. 
Upon further cooling, Josephson-type weak links are formed among the grains.
Therefore, shielding currents flow in inter-grain paths and magnetic flux is excluded from the inter-grain regions.
As a result, two transitions, which are attributed to intra-grain and inter-grain superconductivity, are observed in $\chi_\mathrm{AC}(T)$ and 
the transition temperature of inter-grain superconductivity is sensitive to $H_\mathrm{AC}$ than to $H_\mathrm{DC}$. \cite{Yang1994PRB} 
These features are observed in our results.
Therefore, we first discuss the scenario that small superconducting Sr$_2$RuO$_4$ grains are embedded in the Sr$_3$Ru$_2$O$_7$ region of the eutectic crystals and
superconducting networks are formed among them along $ab$ planes (Fig.~\ref{model} left; Scenario~I).

Here, let us introduce a model developed by M\"uller and Yang \textit{et al.} in order to calculate $\chi_\mathrm{AC}$ of granular superconductors.
Yang \textit{et al.} \cite{Yang1994PRB} calculated $\chi_\mathrm{AC}(T)$ of polycrystal Tl$_2$Ba$_2$Ca$_2$Cu$_3$O$_y$ by a method similar to M\"uller's theoretical work, \cite{Muller1989PhysicaC} and 
their results well reproduced the experimental findings.
Below, we calculate $\chi^\prime$ and $\chi^{\prime \prime}$ in the same way as Yang \textit{et al.} \cite{Yang1994PRB}
The sample shape is assumed to be a thin slab of thickness $2d$ in the $x$ direction. 
The length in the $y$ direction and height in the $z$ direction of the sample are assumed to be infinity.
The applied field $H_\mathrm{a}(t)=\sqrt{2}H_\mathrm{AC}\cos(\omega t)+H_\mathrm{DC}$ is parallel to the slab's $z$ direction. 
To avoid the complication of demagnetization factors, the grains are approximated as 
infinitely-long superconducting cylinders aligned along the $z$ direction. 
Instead, effects of finite demagnetization factors are embedded in other parameters as we will explain below.
The grain radius is assumed to be the same value $R_\mathrm{g}$ ($\ll 2d$), 
which represents the average grain radius in the experiments, for the all grains. 

The real and imaginary parts of $\chi_\mathrm{AC}$ are expressed as
\begin{align}
\chi^{\prime} &= \frac{\omega}{\sqrt{2}\pi \mu_0H_\mathrm{AC}}\int^{2\pi/\omega}_0 \langle B(t) \rangle \cos(\omega t)\mathrm{d}t - 1, \label{chi1} \\
\chi^{\prime\prime} &= \frac{\omega}{\sqrt{2}\pi \mu_0H_\mathrm{AC}}\int^{2\pi/\omega}_0 \langle B(t) \rangle\sin(\omega t)\mathrm{d}t. \label{chi2}
\end{align}
Here, $\langle B(t) \rangle$ is the spatial average local flux density over the sample cross-section, and is given by
$\langle B(t) \rangle =\langle B_\mathrm{J}(t)\rangle_x +\langle\!\langle B_\mathrm{g}(t)\rangle\!\rangle_{r,x}$. 
$\langle B_\mathrm{J}(t)\rangle_x$ and $\langle\!\langle B_\mathrm{g}(t)\rangle\!\rangle_{r,x}$ are 
the spatial average over the sample of the inter-grain flux density $B_\mathrm{J}(x,t)$ and 
that of the average intra-grain flux density threading a cylindrical grain $\langle B_\mathrm{g}(x,t)\rangle_r$, respectively. 
These notations are the same as those given by M\"uller. \cite{Muller1989PhysicaC}

Let us now derive the inter- and intra-grain magnetic field distribution using the critical state equations \cite{Muller1989PhysicaC} 
in order to obtain $\langle B_\mathrm{J}(t)\rangle_x$ and $\langle\!\langle B_\mathrm{g}(t)\rangle\!\rangle_{r,x}$.
For the inter-grain regions, magnetic flux density $B_\mathrm{J}$ is larger than $\mu_0H_\mathrm{J}$ 
because magnetic flux is compressed into the inter-grain regions due to the diamagnetism of the superconducting grains.
By embedding this effect into the effective permeability $\mu_\mathrm{eff}$, 
$B_\mathrm{J}$ can be expressed as $B_\mathrm{J}=\mu_\mathrm{eff}\mu_0H_\mathrm{J}$. 
The effective permeability $\mu_\mathrm{eff}(T)$ is written as \cite{Clem1988PhysicaC,Muller1989PhysicaC-2,Raboutou1980PRL} 
$\mu_\mathrm{eff}(T)=f_\mathrm{n}+f_\mathrm{s}F(R_\mathrm{g}/\lambda_\mathrm{g}(T))$, 
where $\lambda_\mathrm{g}(T)$ denotes the London penetration depth of the superconducting grains, which depends on $T$ as 
$\lambda_\mathrm{g}(T)=\lambda_\mathrm{g}(0)[1-(T/T_\mathrm{cg})^4 ]^{-1/2}$. 
The factor $f_\mathrm{s}$ is the area fraction of the projection of grains onto a plane normal to the magnetic field, and  $f_\mathrm{n}$ is that of inter-grain regions ($f_\mathrm{n}=1-f_\mathrm{s}$).
The flux penetration within the surface penetration depth of the grains in the Meissner state is taken into account via $F(x)$, 
which is written as $F(x)=2I_1(x)/(xI_0(x))$; 
$I_0$ and $I_1$ are the modified Bessel functions of the first kind.
The inter-grain magnetic field distribution $H_\mathrm{J}$ is given by the solution of the critical state equations \cite{Muller1989PhysicaC} 
\begin{equation}
J_\mathrm{cJ}(x,t)=\frac{\alpha_\mathrm{J}(T)}{\mu_\mathrm{eff}(T)\mu_0}\frac{1}{|H_\mathrm{J}(x,t)|+H_\mathrm{0J}}, \label{cr1}
\end{equation}
\begin{equation}
\frac{\mathrm{d}H_\mathrm{J}(x,t)}{\mathrm{d}x}=\pm J_\mathrm{cJ}(x,t), \label{cr2}
\end{equation}
where we assume that the pinning force $\alpha_\mathrm{J}$ of a vortex is equal to the Lorentz force. \cite{Tinkham}
The pinning force $\alpha_\mathrm{J}$ is assumed to depend on $T$ as \cite{Yang1994PRB}
$\alpha_\mathrm{J}(T)/\mu_\mathrm{eff}(T) = \alpha_\mathrm{J}(0)(1-T/T_\mathrm{cJ})^2/\mu_\mathrm{eff}(0)$, and
$H_\mathrm{0J}$ is a positive parameter.
The $\pm$ signs account for the outward or inward motion of vortices with decreasing or increasing applied magnetic field, respectively.

While for the intra-grain regions, we also assume the critical state.
Here, we used $B_\mathrm{g}(r,x,t)$, which is equal to $\mu_0(H^\mathrm{ext}_\mathrm{g}(x,t)+M_\mathrm{g}(r,x,t))$, where
$H^\mathrm{ext}_\mathrm{g}$ is the magnetic field at the boundary of a grain and $M_\mathrm{g}$ is the local magnetization in the grain,
because $M_\mathrm{g}$ is finite in the grains.
$B_\mathrm{g}(r,x,t)$, which is equivalent to $\mu_0H_\mathrm{g}(r,x,t)$ in M\"uller's work, \cite{Muller1989PhysicaC} 
is obtained by the solution of the equations 
\begin{equation}
J_\mathrm{cg}(r,x,t)=\frac{\alpha_\mathrm{g}(T)}{|B_\mathrm{g}(r,x,t)|+B_\mathrm{0g}}, \label{cr3}
\end{equation}
\begin{equation}
\frac{1}{\mu_0}\frac{\mathrm{d}B_\mathrm{g}(r,x,t)}{\mathrm{d}r}=\pm J_\mathrm{cg}(r,x,t). \label{cr4}
\end{equation}
In this case, the pinning force $\alpha_\mathrm{g}$ is assumed \cite{Muller1989PhysicaC} to be $\alpha_\mathrm{g}(T) = \alpha_\mathrm{g}(0)[1-(T/T_\mathrm{cg})^2]^2$. 
$B_\mathrm{0g}$ is a positive parameter.
We note that the effects of demagnetization factors are embedded in $\alpha_\mathrm{g}(0)$ and $B_\mathrm{0g}$.
For example, if the demagnetization factor is large, $\alpha_\mathrm{g}(0)/B_\mathrm{0g}$ would become large.
Solving Eqs.~(\ref{cr1})\,--(\ref{cr4}), we obtain $H_\mathrm{J}(x,t)$ and $B_\mathrm{g}(r,x,t)$, 
from which we can calculate $\langle B_\mathrm{J}(t)\rangle_x$ and $\langle\!\langle B_\mathrm{g}(t)\rangle\!\rangle_{r,x}$, 
as shown in Ref.~\onlinecite{Muller1989PhysicaC}. 
By putting these quantities into Eqs.~(\ref{chi1}) and (\ref{chi2}),
we obtain $\chi^\prime$ and $\chi^{\prime\prime}$. 
Hereafter, we call this model the M\"uller-Yang model.

In this calculation, we fixed $T_\mathrm{cg}$, $T_\mathrm{cJ}$, and $d$ to the values obtained from the present measurements, 
i.~e. $T_\mathrm{cg}=1.34$~K, $T_\mathrm{cJ}=1.10$~K, and $d=1$ mm, and
$\lambda_\mathrm{g}(0)$ to the known value for the bulk Sr$_2$RuO$_4$.
We varied the other parameters so that the calculated results best agree with our experiments: 
$f_\mathrm{n}$ and $R_\mathrm{g}$ are changed manually so that we reproduce the behavior observed in weak magnetic fields at temperatures around $T_\mathrm{cg}$, 
$H_\mathrm{0J}$ and $\alpha_\mathrm{0J}$ are adjusted so that we reproduce the AC magnetic field dependence of the step in $\chi^\prime$ and the peak in $\chi^{\prime\prime}$ at transition~C, 
and $B_\mathrm{0g}$ and $\alpha_\mathrm{0g}$ are adjusted so that we reproduce the AC magnetic field dependence of transition~B. 

The calculated results based on the M\"uller-Yang model are shown in Figs.~\ref{calc}(c) and (d).
In our calculations, $R_\mathrm{g}$ was estimated to be 2\ $\muup$m, 
and $f_\mathrm{s}$ was estimated to be 60\%.
These parameters are similar to those used in M\"uller's work.\cite{Muller1989PhysicaC}
However, our calculation contains several inconsistences with the experiments.
First, the behavior of transition~C is different from that of the Josephson weak-link network,
as shown in Figs.~\ref{calc}(a)-(d).
In our calculations, $T^*_\mathrm{C}$ is shifted toward lower temperatures, but 
$\Delta \chi^\prime_\mathrm{C}$ is not severely suppressed with increasing the amplitude of $H_\mathrm{AC}$,
which is typical weak-link behavior. 
In contrast, in our experiments, \textit{both} $T^*_\mathrm{C}$ and $\Delta \chi^\prime_\mathrm{C}$ are easily suppressed with increasing the strength of $H_\mathrm{AC}$.
Moreover, as shown in Figs.~5 and 6, the temperature dependences of $H^*_\mathrm{B}$ and of $H^*_\mathrm{C}$ are qualitatively similar, and are different from $H_\mathrm{c2}(T)$ of bulk Sr$_2$RuO$_4$. 
This behavior is not consistent with a model of granular superconductivity, 
in which $T_\mathrm{c}$ of grains and $T_\mathrm{c}$ of inter-grain region should exhibit totally-different field dependences.
These results suggest that the Josephson-network scenario does not seem to be suitable for the superconductivity in the Sr$_3$Ru$_2$O$_7$ region of the eutectic crystals.

\subsection*{Scenario II}

The second scenario assumes that no superconducting network is formed, but 
superconductors of thin-film shapes with multiple $T_\mathrm{c}$'s are contained in the Sr$_3$Ru$_2$O$_7$ region. 
We consider that stacked monolayers of RuO$_2$ planes, the building block of superconducting Sr$_2$RuO$_4$, 
are contained in the Sr$_3$Ru$_2$O$_7$ region as stacking faults, and 
exhibit superconductivity with different $T_\mathrm{c}$'s depending on the number of monolayers contained in a stacking unit. 
Although fabrications of superconducting thin films of Sr$_2$RuO$_4$ have not been reported so far, 
it was reported that $T_\mathrm{c}$ of thin YBa$_2$Cu$_3$O$_{7-x}$ films depends on their thickness.\cite{Varela2002PRB}
It is also known that the $H_\mathrm{c2}(T)$ curve of quasi-two-dimensional superconductors for $H_\mathrm{DC} \perp$~layer, \cite{Taniguchi1998PRB} 
which can be regarded as a stacking of thin films, has a positive curvature near $H_\mathrm{DC}=0$.
The thickness of monolayers should be comparable to or less than the coherence length of Sr$_2$RuO$_4$ along the $c$ axis ($\sim$~3.3~nm) 
because the transitions B and C would behave as bulk superconductivity if the thickness were much larger than the coherence length.
This scenario is consistent with the fact that we cannot find Sr$_2$RuO$_4$ in the Sr$_3$Ru$_2$O$_7$ region by EDX analysis and PLOM images 
because Sr$_2$RuO$_4$ slabs with a thickness of several nano meters are too thin to find for our instruments.

Although this scenario appears to be different from the situation of Scenario I, 
we can still calculate $\chi_\mathrm{AC}(T)$ using the M\"uller-Yang model after a slight modification. 
The modified model, which we call as a multiple superconductor model,
assumes that the sample is divided into areas with different $T_\mathrm{c}$'s and 
$\chi_\mathrm{AC}(T)$ is calculated in each area using Eqs.~(\ref{chi1})-(\ref{cr4}) with $\alpha_\mathrm{J}(0)=0\ \mathrm{TAm^{-2}}$ and $f_\mathrm{n}=0$. 
These conditions assume that no superconducting network is formed in the sample.
In this model calculation, we considered the AC susceptibility of the sample as $\chi_\mathrm{AC}=\sum_{i} p_i\, \chi_i$, 
where $p_i$ is the percentage of the $i$-th area ($\sum_i p_i=1$), and 
$\chi_i$ represents the AC susceptibility of the $i$-th area. 
It might be more plausible in reality that the thickness of a single Sr$_2$RuO$_4$ thin slab is not homogeneous. 
This possible inhomogeneity was neglected in our calculations.

For Sample~1b, we assumed three kinds of superconductors SC1, SC2, and SC3, with two distinct transition temperatures
to reproduce the experiments well.
The necessity of introducing SC3 implies that 
there are two kinds of regions with essentially the same $T_\mathrm{c}$, but with much different $J_\mathrm{c}$ values.
These different $J_\mathrm{c}$ values would be caused by the effects of finite demagnetization factor of thin film 
because we embedded it into the parameters $B_\mathrm{0g}$ and $\alpha_\mathrm{g}(0)$.
However, we consider that the existence of SC3 is not essential
because SC3 was not necessary in the calculations for other samples.

\begin{figure}
\includegraphics[width=3.3in]{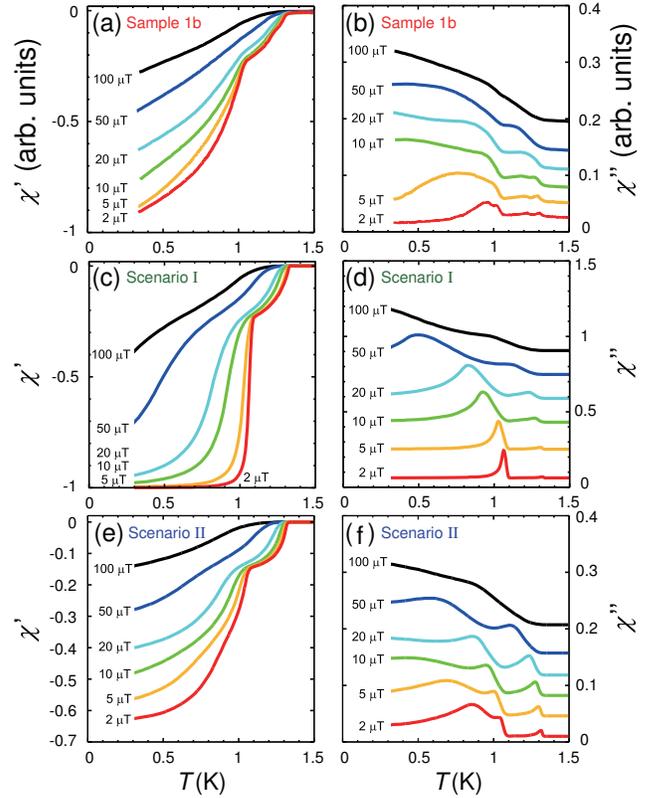}
\caption{(Color online) (a), (b) Experimental and (c)-(f) calculated results of $\chi_\mathrm{AC}(T)$ at various AC magnetic fields in $\mu_0H_\mathrm{DC}=0$ T.
The numbers labeling the curves indicate the applied AC field amplitude $\mu_0H_\mathrm{AC}$ in $\muup$T-rms.}
\label{calc}
\end{figure}

In our calculation, the parameters $T_\mathrm{cg}$, $\lambda_\mathrm{g}(0)$, and $p_i$ were fixed. 
The other parameters $B_\mathrm{0g}$, $\alpha_\mathrm{g}$, and $R_\mathrm{g}$ were adjusted manually 
so that the calculated results best agree with our experiments.
The results are summarized in Table~\ref{para} and Figs.~\ref{calc}(e) and (f). 
Our calculation reproduces the essential features of the experimental findings.
For example, the observation that 
\textit{both} $T^*_\mathrm{C}$ and $\Delta \chi^\prime_\mathrm{C}$ decrease with increasing the amplitude of $H_\mathrm{AC}$ is reproduced. 
Although the critical current density $J_\mathrm{c}(0)$ of the pure Sr$_2$RuO$_4$ is approximately $500\ \mathrm{A/cm^2}$ (Ref.~\onlinecite{Deguchi}), 
that of SC1 was estimated to be $3\times10^5\ \mathrm{A/cm^2}$ from $B_\mathrm{0g}$ and $\alpha_\mathrm{g}(0)$ using Eq.~(\ref{cr3}).
If the thickness of the superconductor decreases, 
the cross section of the superconductor also decreases and the critical current density should become large as often observed in thin films.
Therefore, such a large $J_\mathrm{c}(0)$ may also support the scenario that Sr$_2$RuO$_4$ is contained as a thin slab.

\begin{table}
\begin{center}
\caption{Parameters used in our calculations based on Scenario II. 
The open squares ($\square$) mark the parameters that we adjusted to obtain the best fit to our experiments. 
The closed squares ($\blacksquare$) label the parameters we fixed in our calculations. 
``SC1'' denotes the superconducting area responsible to the transition~B and
``SC2'' and ``SC3'' denote those to the transition~C.
}
\renewcommand{\arraystretch}{1.1}
\begin{tabular*}{8cm}{@{\extracolsep{\fill}}cccccc}\hline
 & & SC1 & SC2 & SC3 &  \\ \hline
$T_\mathrm{cg}$ & (K) & 1.34 & 1.10 & 1.10 & $\blacksquare$ \\
$B_\mathrm{0g}$ & ($\muup$T) & 10 & 30 & 200 & $\square$ \\
$\alpha_\mathrm{g}(0)$ & ($\mathrm{TAm^{-2}}$) & 30000 & 8000 & 3000 & $\square$ \\
$R_\mathrm{g}$ & ($\muup$m) & 1 & 1 & 1 & $\square$ \\
$\lambda_\mathrm{g}$(0) & ($\muup$m) & 0.18 & 0.18 & 0.18 & $\blacksquare$ \\
$p_i$ & & 0.25 & 0.35 & 0.4 & $\blacksquare$ \\ \hline
\end{tabular*}
\label{para}
\end{center}
\end{table}

No single plane of a monolayer RuO$_2$ probably covers the whole $ab$ plane. 
However, the magnetic flux would be excluded from the whole sample for $H \parallel c$ if there are many such layers in the sample.
In addition, this scenario does not contradict with our results that the apparent shielding fraction is less than 1\% for $H \parallel ab$.

From the discussions above, we consider that this scenario is the most probable one to explain the superconductivity 
observed in the Sr$_3$Ru$_2$O$_7$ region of eutectic crystals.
In addition, recently such stacked monolayers of RuO$_2$ planes have indeed been observed using a transmission electron microscope. \cite{Fittipaldi}

\subsection*{The possibility of superconducting Sr$_3$Ru$_2$O$_7$}

Finally, we discuss the possibility that small Sr$_3$Ru$_2$O$_7$ parts in the Sample~1b become superconducting, 
due to a specific arrangement of the RuO$_6$ octahedra, different from the arrangement realized in bulk Sr$_3$Ru$_2$O$_7$.
The structure of bulk Sr$_3$Ru$_2$O$_7$ contains orthorhombic deformations due to the rotation of the RuO$_6$ octahedra. \cite{Shaked2000PRB,Huang1998PRB,Inoue1997}
In Ruddlesden-Popper type ruthenates Sr$_{n+1}$Ru$_n$O$_{3n+1}$, 
it is known that their rotation, tilting, and flattening of RuO$_6$ octahedra affect the electronic states significantly. \cite{Matzdorf2000Science,Friedt2001PRB,Fang2001PRB} 
In fact, the electronic and thermodynamic properties of Ca$_{2-x}$Sr$_x$RuO$_4$ are greatly affected by 
the rotation with varying $x$ (Refs.~\onlinecite{Fang2001PRB}, \onlinecite{Kriener2005PRL}).
Degrees of freedom such as the rotation angle or an ordering pattern of rotations might be left in Sr$_3$Ru$_2$O$_7$ under certain circumstances.
Indeed, different ordering patterns of rotations have been reported in powder samples. \cite{Shaked2000PRB,Huang1998PRB,Inoue1997}
Therefore, it is possible that some small parts of Sr$_3$Ru$_2$O$_7$ with a certain arrangement of RuO$_6$ octahedra are superconducting and 
that these parts play roles of superconductors in Scenario II.
However, we did not obtain any direct structural evidence to conclude that octahedral rotation and/or tilting in eutectic Sr$_3$Ru$_2$O$_7$ is different from that in bulk Sr$_3$Ru$_2$O$_7$.
On the basis of the available information, therefore, so far we cannot conclude that Sr$_3$Ru$_2$O$_7$ itself is superconducting.

\section{Summary}
We have studied superconductivity in the Sr$_3$Ru$_2$O$_7$\,--\,Sr$_2$RuO$_4$ eutectic system.
Our AC susceptibility measurements revealed that 
multiple superconducting transitions occur in the Sr$_3$Ru$_2$O$_7$\,--\,Sr$_2$RuO$_4$ eutectic sample, and 
that the transitions with $T_\mathrm{c}$ lower than that of Sr$_2$RuO$_4$ originate from the Sr$_3$Ru$_2$O$_7$ region alone.
These experimental results indicate that the superconductivity observed in the Sr$_3$Ru$_2$O$_7$ region is not 
attributable to an unusually-long-range proximity effect across the boundary between Sr$_2$RuO$_4$ and Sr$_3$Ru$_2$O$_7$. 
Both $T^*$ and $\Delta \chi^\prime$ of this superconductivity are sensibly suppressed by weak AC magnetic fields.
Moreover, their $H$\,--$T$ phase diagrams are qualitatively different from that of bulk Sr$_2$RuO$_4$, 
and no anomaly was observed in the specific heat of the Sr$_3$Ru$_2$O$_7$ region sample cut from the eutectic crystals.
Although we have not achieved a conclusive explanation of the origin of superconductivity in the Sr$_3$Ru$_2$O$_7$ region, 
we proposed scenarios to explain our experiments. 
Among them, the scenario in which Sr$_2$RuO$_4$ thin slabs are embedded in the Sr$_3$Ru$_2$O$_7$ region and 
the multiple superconducting transition temperatures arise from the distribution of the slab thickness 
yielded the most satisfying fit to the experiments.

\acknowledgments
We thank Franceso Tafuri, Manfred Sigrist, Mario Cuoco, Canio Noce, Yukio Tanaka, Yasuhiro Asano, Robin S. Perry, and Andrew P. Mackenzie for valuable discussions, and
Kentaro Kitagawa, Hiroshi Takatsu, and Markus Kriener for their support. 
This work has been supported by Grants-in-Aid for Scientific Research from the Japan Society for Promotion of Science (JSPS) and
from the Ministry of Education, Culture, Sports, Science, and Technology of Japan (MEXT). 
It is also supported by a Grant-in-Aid for the 21st Century COE program ``Center for Diversity and Universality in Physics'' from MEXT.

\end{document}